\documentstyle[11pt,aaspp4]{article}

\begin{document}

\title{Detecting Scale-Dependence of Bias from APM-BGC Galaxies}

\author{Li-Zhi Fang$^{1}$, 
Zu-Gan Deng$^{2,4}$ 
and Xiao-Yang Xia$^{3,4}$ 
} 
\affil{$^{1}$Department of Physics, University of Arizona, Tucson, AZ 85721}
\affil{$^2$Department of Physics, Graduate School, USTC, Beijing 100039, China}
\affil{$^3$Department of Physics, Tianjin Normal University, 
300074 Tianjin,  China}
\affil{$^4$Beijing Astronomical Center, Chinese Academy of Sciences and Peking
University, Beijing 100080, China}

\begin{abstract}
We present an investigation of the scale-dependence of bias described by
the linear model: $(\delta \rho({\bf x})/\bar{\rho})_{g} = 
b (\delta \rho({\bf x})/\bar{\rho})_{m}$, $b$ being the bias parameter, and
 $\rho({\bf x})_{g}$ and  $\rho({\bf x})_{m}$ are the galaxy number density 
and mass density, respectively. Using a discrete wavelet decomposition, we
show that the behavior of bias scale-dependence cannot be described 
by one parameter $b$. In the linear bias model the scale-dependence should be 
measured by the $j$-spectra of wavelet-coefficient-represented bias parameters  
$\tilde{b}^{(n)}_j$ and $b_j^{(n)}$, $n$ being  positive integers. Because 
$\tilde{b}^{(n)}_j$ with different $n$ are independent from each other, 
a systematic analysis of the $j$-spectra of $\tilde{b}^{(n)}_j$ and 
$b_j^{(n)}$ is necessary.

We performed a $j$-spectrum analysis for samples of elliptical and 
lenticular (EL), and spiral (SP) galaxies listed in the APM bright galaxy 
catalog. We found that, for statistics of two-point correlation functions or
DWT power spectrum, the scale-independence holds within 1 $\sigma$. However, 
the bias scale-dependence becomes substantial when phase-sensitive statistics 
(e.g. $\tilde{b}^{(n)}_j$ with $n>2$ or $b_j^{(n)}$) are applied. These results
indicate that the bias
scale-dependence has the same origin as the non-Gaussianity of galaxy
distributions. This is generally consistent with the explanation that the
bias scale-dependence originated from non-linear and non-local
relationship between galaxy formation and their environment.
\end{abstract}

\bigskip

\keywords{cosmology: theory - galaxies: statistics - large scale structure}

\newpage

\section{Introduction}

Bias is introduced to reconcile the amplitude of fluctuations inferred 
from clustering of galaxies with that derived from mass distributions.
Bright galaxies seem to have a stronger clustering  than that of the
underlying mass. Therefore, it is generally believed that galaxies are
biased tracers of the mass density field, i.e. it follows the clustering 
of (dark) matter but with an enhanced amplitude.

Generally, bias is phenomenologically modeled by a linear relation as
\begin{equation}
\delta({\bf x})_{g} = b \delta({\bf x})_{m}.
\end{equation}
where $\delta({\bf x})_{g} = [n({\bf x})-\bar{n}]/\bar{n}$, 
$\delta({\bf x})_{m} = [\rho({\bf x})-\bar{\rho}]/\bar{\rho}$, 
$\rho({\bf x})$ and $n({\bf x})$ are, respectively, the galaxy 
number density distribution and  density field of dark matter, and
$\bar{\rho}$ and $\bar{n}$ being the average of $\rho({\bf x})$ and 
$n({\bf x})$. The phenomenological bias parameter $b$ is assumed to be 
constant, but may be different for different galaxy
types, say, $b_{early}$, $b_{late}$ for early and late types of galaxies.  

Eq.(1) and its variants are widely used in the determination of 
cosmological parameters from samples of redshift surveys of galaxies. However,
we have really very little idea about which the bias parameter $b$ should be.
In fact, both bias ($b>1$) and anti-bias ($b<1$) are employed in current data 
analysis (e.g. Mo, Jing \& White 1996.) This prevents unambiguous measures
of cosmological parameters, giving only bias-contaminated results. 

Theoretically, the physical mechanism responsible for relation (1) is 
far from clear. The first analytic model of bias, in which objects 
are identified with high peaks or collapsed halos of the density field, 
offers a plausible explanation of the bias of galaxy clusters -- the 
correlation amplitudes of clusters are strong functions of cluster 
richness (Kaiser 1984). In this case, bias is mainly caused by the mass of
collapsed halos. The larger the mass of the halo, the higher the richness 
of the cluster. 

Yet, the formation of galaxies doesn't depend only on the mass,
or local mass density, of collapsed halos, but is substantially modulated 
by various environmental effects, such as the suppression of star formation 
in neighboring protogalaxies (Rees 1985), stimulating the formation of nearby
galaxies (Dekel \& Rees 1987), dynamical friction effects
(Couchman \& Carlberg 1992),
etc. All these environmental effects are beyond local density, and lead to 
a non-local relation between the number density of galaxies 
and the background mass field (Bower et al. 1993). Moreover, the rate of star
formation is most likely non-linearly dependent on local mass density.
A common result of the non-local and/or non-linear relation between
$\delta_{g}$ and $\delta_{m}$ is the scale-dependence of parameter $b$.
Therefore, to have a deep 
understanding of the mechanism of galaxy bias, searching for the
scale-dependence of parameter $b$ is necessary  (Coles 1993, Catelan et al. 
1994.)

So far, the results of detecting $b$ scale-dependence are quite scattered.
 For instance, the values of the bias-contaminated density parameter
$\beta=\Omega^{0.6}/b$
are found to be in the range of 0.4 - 1, and equal to about $\sim 0.5-0.6$ 
at Gaussian smoothing scales of 3-6 h$^{-1}$Mpc, and $\sim 1$ on scales of
$\sim 12$ h$^{-1}$Mpc (e.g. Dekel, Burstein \& White 1996). This is, $b$ is 
probably scale-dependent from 6 to 12 h$^{-1}$ Mpc. On the other hand, some 
studies conclude that for galaxies of all types and luminosities the scale
dependence of the bias parameter is weak
(e.g. Kauffmann, Nusser \& Steinmetz 1997). Why different detections gave 
different conclusions? This question motivated us to study the physics 
included in Eq.(1). 

In the first part of this paper, we show that even in linear
bias model, the behavior of scale-dependence cannot be described by one 
parameter $b$, but by a series of $j$-spectra of wavelet-coefficient-represented 
bias parameters $\tilde{b}^{(n)}_j$ and $b_j^{(n)}$, $n$ being  positive 
integers. The $j$-spectra of $\tilde{b}^{(n)}_j$ are all statistically 
{\it independent}. {\it Different} detections may actually measure 
{\it different} parameters $\tilde{b}_j^{(n)}$. 
Therefore, it should not be surprised that some detections are positive, and 
some negative. The different behavior of bias scale-dependence given by different 
detection does not cause confusion, but may greatly be helpful to reveal the
physics behind the bias model (1). Therefore, to have a complete picture of bias 
scale-dependence a systematic analysis of the $j$-spectra of
$\tilde{b}^{(n)}_j$ and $b_j^{(n)}$ is necessary.

In the second part, we performed a systematic detection of the bias 
scale-dependence with the samples of 
galaxies listed in the APM bright galaxies catalog (APM-BGC). This analysis 
shows that for the APM-BGC sample, the scale-independence 
approximately holds if only the two-point correlation function and power
spectrum are involved, while the scale-dependence becomes substantial
when higher order or phase-sensitive statistics involved. This result is 
worth to constrain models of hierarchical clustering of galaxies.

The paper is organized as follows. In \S 2 we briefly introduce a method 
of the space-scale decomposition based on discrete wavelet transform (DWT)
analysis. With this method, various statistics of measuring the bias scale
dependence are developed and presented. \S 3 describes the sample to be
analyzed. In \S 4 we apply the DWT method to analyze the galaxy sample of
the APM bright galaxies. The implications of these results are discussed in 
\S 5.

\section{Linear bias model in DWT representation}

\subsection{The discrete wavelet transform}

Let us briefly introduce the DWT analysis of large scale structures, for 
the details referring to (Fang \& Pando 1997, Pando \& Fang 1996, 1998.)
We consider here a 1-D mass density distribution  $\rho(x)$ 
or contrast $\delta(x)=[\rho(x)-\bar{\rho}]/\bar{\rho}$, which are 
mathematically random fields over a spatial range $0 \leq x \leq L$. It 
is not difficult to extend all results developed in this section into 
2-D and 3-D because the DWT bases for 
higher dimension can be constructed by a direct product of 1-D bases.

Like the Fourier expansion of the field $\delta(x)$, the DWT expansion of 
the field $\delta(x)$ is given by
\begin{equation}
\delta(x) = \sum_{j=0}^{\infty} \sum_{l= 0}^{2^j -1}
  \tilde{w}_{j,l} \psi_{j,l}(x)
\end{equation}
where $\psi_{j,l}(x)$, $j=0,1,...$, $l=0...2^j-1$ are the bases of the DWT.
Because these bases are orthogonal and complete, the wavelet function
coefficient (WFC), $\tilde{w}_{j,l}$, is computed by
\begin{equation}
\tilde{w}_{j,l}= \int \delta(x) \psi_{j,l}(x)dx.
\end{equation}

The wavelet transform bases $\psi_{j,l}(x)$ are generated from the 
basic wavelet $\psi(x/L)$ by a dilation $2^j$, and a translation 
$l$, i.e. 
\begin{equation}
\psi_{j,l}(x) =\left( \frac{2^j}{L} \right)^{1/2} \psi(2^jx/L-l).
\end{equation}
The basic wavelet $\psi$ is designed to be continuous, admissible 
and localized. Unlike the Fourier bases $\exp (i2\pi nx/L)$, which are 
non-local in 
physical space, the wavelet bases $\psi_{j,l}(x)$ are localized in 
both physical space and  Fourier (scale) space. In physical space, 
$\psi_{j,l}(x)$ is centered at position $lL/2^j$, and in Fourier 
space, it is centered at wavenumber $2\pi \times 2^j/L$. Therefore,
the DWT decomposes the density fluctuating field 
$\delta(x)$ into domains ${j,l}$ in phase space, and for each basis the
corresponding area in the phase space is as small as that allowed by the
uncertainty principle. WFC $\tilde{w}_{j,l}$ and its intensity
$|\tilde{w}_{j,l}|^2$ describe, respectively,
the fluctuation of density and its power on scale $L/2^j$ at position
$lL/2^j$.

In order to have reasonable statistics, the cosmic density field 
is usually assumed to be ergodic: the average over an ensemble is equal to 
the spatial average taken over one realization. This is the so-called 
``fair sample hypothesis" (Peebles 1980). A homogeneous
Gaussian field with continuous spectrum is certainly ergodic (Adler 1981).
In some non-Gaussian cases, such as homogeneous and isotropic turbulence
(Vanmarke, 1983), ergodicity also approximately holds. Roughly, the
ergodic hypothesis is reasonable if spatial correlations are decreasing
sufficiently rapidly with increasing separation.  The volumes separated
with distances larger than the correlation length are approximately
statistically independent. In this case, WFCs, $\tilde{w}_{j,l}$, 
at different $l$ from one realization of $\delta(x)$ can be employed as
statistically independent measurement. Thus, the WFCs of 
$\delta(x)$ at different $l$ on a given $j$, $\tilde{w}_{j,l}$, form an 
ensemble of the WFCs on the scale $j$. In other words, when the ``fair 
sample hypothesis" holds, an 
average  over ensemble can be fairly estimated  by average over $l$, i.e.
$\langle \tilde{w}_{j,l} \rangle \simeq
(1/2^j)\sum_{l=0}^{2^j-1}\tilde{w}_{j,l}$,
where $\langle ...\rangle$ denotes ensemble average.
 
\subsection{Reconstruction of density field}

Using the completeness of the DWT basis, one can reconstruct the original
density field from the coefficient $\tilde{w}_{j,l}$. To
achieve this, DWT analysis employs another set of functions consisting of
the so-called scaling functions, $\phi_{j,l}$, which are generated from
the basic scaling $\phi(x/L)$ by a dilation $2^j$, and a translation $l$,
i.e.
\begin{equation}
\phi_{j,l}(x) = \left( \frac{2^j}{L} \right )^{1/2} \phi(2^jx/L - l).
\end{equation}
The basic scaling $\phi$ is essentially a window function with width
$x/L=1$. Thus, the scaling functions $\phi_{j,l}(x)$ are also windows, 
but with width $(1/2^j)L$, and centered at $lL/2^j$.
The scaling functions $\phi_{j,l}(x)$ are orthogonal with respect to 
the index $l$, but not for $j$. This is a common property of window 
functions, which 
can be orthogonal in physical space, but not in Fourier space. 

For Daubechies wavelets, the basic wavelet and the basic scaling are
related by recursive equations as (Daubechies 1992) 
\begin{equation}
\begin{array}{ll}
\phi(x/L) & = \sum_l a_{l} \phi(2x/L-l) \\
\psi(x/L) & = \sum_{l} b_{l} \phi(2x/L + l)
\end{array}
\end{equation}
where coefficients $a_l$ and $b_l$ are different for different wavelet. 
In this paper, we use the Daubechies 4 wavelet (D4), for which
$a_0=(1+\sqrt 3)/4, \ a_1=(3+\sqrt 3)/4, \ a_2=(3-\sqrt 3)/4, \
a_3=(1-\sqrt 3)/4$. 

From Eq.(6), one can show that the scaling functions $\phi_{j,l}(x)$ are
always orthogonal to the wavelet bases $\psi_{j',l'}(x)$ if $j \leq j'$, i.e.
\begin{equation}
\int \phi_{j,l}(x)\psi_{j',l'}(x)dx = 0, 
\hspace{2cm} {\rm for \ \ \ } j \leq j'.
\end{equation}
Therefore, $\phi_{j,l}(x)$ can be expressed by
$\psi_{j',l'}(x)$ as
\begin{equation}
\phi_{j,l}(x) = \sum_{j'=0}^{\infty}\sum_{l'=0}^{2^{j'}-1}
c_{jl;j'l'} \psi_{j',l'}(x) =
\sum_{j'=0}^{j-1}\sum_{l'=0}^{2^{j'}-1}
c_{jl;j'l'} \psi_{j',l'}(x).
\end{equation}
The coefficients $c_{jl;j'l'}= \int \phi_{j,l}(x)\psi_{j',l'}(x)dx$
can be determined from $a_l$ and $b_l$.

Using $\phi_{j,l}(x)$, we construct a density field on
scale $j$ as
\begin{equation}
\rho^{j}(x) = \sum_{l=0}^{2^{j}-1} w_{j,l} \phi_{j,l}(x),
\end{equation}
where $w_{j,l}$ is called the scaling function coefficient (SFC) given by
\begin{equation}
w_{j,l}= \int_0^{L} \rho(x) \phi_{j,l}(x)dx.
\end{equation}
Since the scaling function 
$\phi_{j,l}(x)$ is window-like, the coefficient $w_{j,l}$ is actually a 
``count-in-cell" in a window on scale $j$ at position $l$.

Using Eqs.(2), (8), (9) and (10), one can find
\begin{equation}
\rho^j(x) = \bar{\rho}\sum_{j'=0}^{j-1} \sum_{l'= 0}^{2^{j'} -1}
  \tilde{w}_{j',l'} \psi_{j',l'}(x) + \bar{\rho}.
\end{equation}
Namely, $\rho^j(x)$ contains all terms of density fluctuations
$\tilde{w}_{j',l'}\psi_{j',l'}(x)$ of $j' <j$, but not terms 
of $j' \geq j$.  From Eqs.(2) and (11), we have
\begin{equation}
\rho(x)=\rho^j(x) + \bar{\rho}
\sum_{j'=j}^{\infty} \sum_{l'= 0}^{2^{j'} -1}
  \tilde{w}_{j',l'} \psi_{j',l'}(x).
\end{equation}
One can also define the smoothed density contrasts on scale $j$ 
to be 
\begin{equation}
\delta^j(x) \equiv \frac{\rho^j(x)-\bar{\rho}}{\bar{\rho}}
= \sum_{j'=0}^{j-1} \sum_{l'= 0}^{2^{j'} -1}
  \tilde{w}_{j',l'} \psi_{j',l'}(x).
\end{equation}
Eqs.(12) and (13) show that $\rho^j(x)$ and $\delta^i(x)$ are the 
density field smoothed on scale $j$. Using Eqs.(12) and (13), one 
can construct the density field
$\rho^j(x)$ or $\delta^j(x)$ on finer and finer scales by WFCs
$\tilde{w}_{j,l}$ till to the precision of the original field. Since the
sets of bases
$\psi_{j,l}$ and $\phi_{j,l}$ are complete, the original field can be 
reconstructed without lost information. 

\subsection{Statistical description of scale dependence of bias}

We subject the DWT decomposition on Eq.(1) which gives
\begin{equation}
(\tilde{w}_{j,l})_{g} = b (\tilde{w}_{j,l})_{m}.
\end{equation}
Statistically, Eq.(14) means that the one-point distributions of the WFCs
for galaxies should be proportional to the matter, and the proportional
coefficient $b$ is $j$-independent. Thus, linear bias models Eq.(1)
requires that various statistics based on the WFC one-point distributions
for galaxies and matter should satisfy the linear relations given by Eq.(14).

If we define the cumulant moment of the one-point distribution 
of $\tilde{w}_{j,l}$ as
\begin{equation}
C^n_j= \langle \tilde{w}_{j,l}^n \rangle = \frac{1}{2^j}
\sum_{l=0}^{2^j-1} \tilde{w}_{j,l}^n, 
\end{equation}
the model (1) is actually to require that the ratios 
$(C^n_j)_{g}/(C^n_j)_{m}=
\langle(\tilde{w}_{j,l})_{g}^n\rangle/ 
\langle (\tilde{w}_{j,l})_{m}^n\rangle$ should be $j$(scale)-independent 
for all order $n$. Namely,
\begin{equation}
\tilde{b}^{(n)}_j = \frac{(C^n_j)_{g}}{(C^n_j)_{m}}
\end{equation}
should be $j$-independent. In other words, the $j$-spectra of 
$\tilde{b}^{(n)}_j$ are flat.
For non-Gaussian fields, the parameters $\tilde{b}^{(n)}_j$ for different $n$
generally are statistically independent. Therefore, the behavior of bias 
scale-dependence cannot be described by one parameter $b$, but the $j$-spectra
of $\tilde{b}^{(n)}_j$ for all $n$.

From Eqs.(13) and (14), one has
\begin{equation}
\delta^j(x)_{g} = b \delta^j(x)_{m}.
\end{equation}
Eq.(17) can be rewritten as
\begin{equation}
\rho^j(x)_{g}-\bar{\rho}_{g}= b' [\rho^j(x)_{m}-\bar{\rho}_{m}]
\end{equation}
where constant $b'=b\bar{\rho}_{g}/\bar{\rho}_{m}$.
 Considering that
$\int \phi_{j,l}(x)dx$ is independent 
of index $l$, the DWT decomposition of Eq.(18) gives
\begin{equation}
(w_{j,l}- \overline{w}_j)_{g} = b'(w_{j,l}-\overline{w}_j)_{m}
\end{equation}
where $\overline{w}_j=(1/2^j)\sum_{l=0}^{2^j-1}w_{j,l}$ is the mean of the SFCs
on scale $j$. Similar to Eq.(16), one can design statistics $b^{(n)}_j$ as follows
\begin{equation}
b^{(n)}_j = \frac{\sum_{l=0}^{2^j-1}(w_{j,l}- \overline{w}_j)^n_{g}}
        {\sum_{l=0}^{2^j-1}(w_{j,l}- \overline{w}_j)^n_{m}}.
\end{equation}
If bias is scale-independent, the $j$-spectra of $b^{(n)}_j$ should also be
flat.

In a word, the scale-dependence of bias is described by the $j$ spectra
of parameters $\tilde{b}^{(n)}_j$ and $b^{(n)}_j$. It is possible that some 
$j$-spectra are flat, but others non-flat. Therefore, to study bias
scale-dependence, it is necessary to take a systematic detection of 
the $j$-spectra of $\tilde{b}^{(n)}_j$ and $b^{(n)}_j$. 

\subsection{Non-Gaussianity and spectra of $\tilde{b}^{(n)}_j$ 
   and $b^{(n)}_j$}

A physical reason of choosing $\tilde{b}^{(n)}_j$ and $b^{(n)}_j$ to
detect scale-dependence of bias is from non-Gaussianity of mass density
fields. 

The structure formation due to gravitational clustering can roughly be 
divided into 
three stages: linear, quasilinear and fully developed nonlinear. It is 
obvious from eq.(1) that the linear evolution of $\delta({\bf x})_{g}$ and 
$\delta({\bf x})_{m}$ do not cause bias scale-dependence. It is generally 
believed that the evolution of cosmic clustering on very large scales is 
probably still remaining in linear stage at present. Therefore, the bias 
parameter should be scale-independent on very large scales if the initial 
$b$ is scale-independent. 

On scales from about 5 - 10 h$^{-1}$ Mpc, the cosmic gravitational 
clustering may also be not fully developed yet, but partially in the 
quasilinear regime. In this case, the power spectrum of density perturbations 
is not 
significantly different from linear power spectrum, but is characterized 
by power transfer via mode coupling. The power transfer among perturbations 
on different scales will give rise to a scale dependence of bias. On the 
other hand, power transfer also leads to non-Gaussianity
of the density fields. Therefore, the bias scale-dependence in the range of
5 - 10 h$^{-1}$ Mpc may not easily be detected by
methods insensitive to non-Gaussianity, such as power spectrum or two-point 
correlation function, but by statistics sensitive to non-Gaussianity. 
The statistics $\tilde{b}^{(n)}_j$ and $b^{(n)}_j$ are non-Gaussian sensitive,
and therefore, suitable to detect the bias scale-dependence caused by
non-Gaussian process.

It has been shown that the $n=2$ statistics of $(\tilde{w}_{j,l})$,
i.e. $\sigma^2_j= C^2_j$, essentially is the Fourier power spectrum. Therefore
a Gaussian random field can be completely described by its DWT power spectrum
$\sigma^2_j= C^2_j$ (Pando \& Fang, 1998.) Higher order ($n>2$) statistics of 
$(\tilde{w}_{j,l})^{n}$ are
sensitive to non-Gaussianity of the density fields. For instance, the
$n=$ 3, 4, statistics of $(\tilde{w}_{j,l})^{n}$ can be rewritten as  
\begin{equation}
S_j \equiv \frac{1}{2^j\sigma^3}C_j^3 \hspace{1cm} 
K_j \equiv \frac{1}{2^j\sigma^4}C_j^4 -3.
\end{equation}
Therefore, $S_j$ and $K_j$ are essentially the same as the statistics of 
skewness and kurtosis. This point can directly be seen from the definition of 
the skewness and kurtosis of density contrast distribution $\delta(x)$. 
They are  
\begin{equation}
S=\frac{1}{\sigma^3}C^3 \hspace{1cm} K=\frac{1}{\sigma^4}C^4-3
\end{equation}
where $C^n=(1/L)\int_0^L\delta^ndx$, and $\sigma^2 = C^2$. Therefore,
$S_j$ and $K_j$ are the $j$-spectra of $S$ and $K$. 

 For Gaussian fields $\delta(x)$, all the cumulant moments $C^n$ or 
$C_j^n$ are equal to zero, only except second order $C^2$ or $C^2_j$
(Fang \& Pando 1997.) Thus, all statistics $\tilde{b}^{(n)}_j$ of $n>2$ 
are good for detecting the scale-dependence of bias related to 
non-Gaussian processes.

The statistics of the SFCs $w_{j,l}$ are different from the WFCs 
$\tilde{w}_{j,l}$. As mentioned in \S 2.3, the 
scaling functions $\phi_{j,l}(x)$ are window-like functions.
SFCs are similar to the statistics of count in cell (CIC).
The DWT scaling functions $\phi_{j,l}(x)$ correspond to the CIC window 
functions, and the SFCs of the DWT correspond to the counts of CIC. Because 
$\phi_{j,l}(x)$ and the window functions of CIC are not orthogonal 
with respect to $j$ (or scale.), the SFCs, $w_{j,l}$, are 
given by a superposition of WFCs, $\tilde{w}_{j',l'}$ and $j' \leq j$
[see, Eqs.(8), (9) and (10)]. 
Namely, for any order of $n$, statistics based on SFCs are always sensitive
to the location, or the phases of perturbations on different scales, i.e. 
sensitive to non-Gaussianity. Thus, regardless whether $n$ is larger than 2,
$b^{(n)}_j$ are always useful to detect the scale-dependence of bias 
related to non-Gaussian processes.

\section{Samples of galaxies}

Neither $\tilde{b}^{(n)}_j$ nor $b^{(n)}_j$ are directly testable
because we don't know the distribution of dark matter $\delta_{m}(x)$.
But, bias scale-dependence can be
revealed from the difference between the scale behaviors of the galaxies 
with different internal properties. It has been very well known that
the relative abundances of galaxies with different morphology are
environment dependent (Dressler 1980), the clustering strength of
galaxies is luminosity dependent (Xia, Deng \& Zhou 1987), and
the correlation amplitudes of optically selected galaxies are different
from that of infrared selected (Saunders, Rowan-Robinson \& Lawrence 1992,
Strauss et al. 1992). Therefore, it is most likely that galaxies with 
different morphology possess different scale-dependence of bias. 
Thus, the bias scale-dependence might be detected by comparing the
$j$-spectra of $\tilde{b}^n_j$ and $b^n_j$ for galaxies with different 
morphology.

Let us consider two types of galaxies, $I$ and $II$, both of which obey
linear model (1) with different parameter $b$. All statistics of Eqs.(15) 
and (20) are available by replacing $(g,m)$ by $(I,II)$. Namely, the scale
dependence of bias can be revealed by $j$-spectrum of $\tilde{b}^{(n)}_j$
and  $b^{(n)}_j $ defined as
\begin{equation}
\tilde{b}^{(n)}_j = \frac{(C^n_j)_{I}}{(C^n_j)_{II}}.
\end{equation}
\begin{equation}
b^{(n)}_j = \frac{\sum_{l=0}^{2^j-1}(w_{j,l}- \overline{w}_j)^n_{I}}
        {\sum_{l=0}^{2^j-1}(w_{j,l}- \overline{w}_j)^n_{II}}.
\end{equation}
 Any non-flatness of these $j$-spectra
indicates that the bias is scale-dependent for at least one type of the
considered galaxies.

We analyzed the samples of galaxies listed in the APM bright galaxies catalog
(Loveday, 1996), which gives positions, magnitudes and morphological types 
of 14,681 galaxies brighter than 16$^m$.44 over a 4,180 deg$^{\circ}$ 
area in 180 Schmidt survey fields of south sky. The completeness is about 
96.3 per cent with a standard deviation 1.9 per cent inferred from carefully 
checked 12 fields. Therefore, it is large and uniform enough for a 2-D DWT 
analysis.

We choose the early type or elliptical and lenticular (EL) galaxies as
$I$, and late type or spiral (SP) galaxies as $II$. All the elliptical and 
lenticular galaxies are compiled into one sample containing 4,439 ELs.
The SP sample contains 8,217 SP galaxies.

In order to do a 2-D DWT analysis, we chose three fields S1, S2 and S3
from the entire survey area. The three fields are selected to be square on
a equal area projection of the sky. The equal area projection keeps the
surface
number density of the galaxies in the plane to be the same as on the sky.
The sizes of S1, S2 and S3 are taken to be as large as possible covering
the whole area of the survey. The whole survey
and the three fields S1, S2 and S3 are plotted in Fig. 1. Each field has
angular size of about 37$^{\circ}\times37^{\circ}$. S1, S2 and S3 contain
1,095, 1,039, and 1,055 ELs, and 2,186, 2,188, and 2,092 SPs, respectively.
The galaxies in the regions S1 and S3 are completely independent. S2 has
some overlaps with S1 and S3.

We divide each square region into 2$^{10}$ $\times$ 2$^{10}$ (1024$^2$) cells
labelled by ${\bf l} =(l_1, l_2)$, and $l_1$, $l_2$ being an integer from $0$ to
1023. Using the estimation of the mean depth of the sample given by the 
luminosity function
from the  Stromlo-APM redshift survey (Loveday et al. 1992), one find that
the cell size is about 65 h$^{-1}$Kpc, which is fine enough to
detect statistical features on scales larger than 1 h$^{-1}$Mpc.

The distribution of the galaxies in the sample can be described by a 2-D 
density field 
$\rho({\bf x})$ or contrast $\delta({\bf x})$, where ${\bf x}=(x_1,x_2)$.
In doing 2-D DWT analysis of $\rho({\bf x})$ or $\delta({\bf x})$, the
wavelet functions and the scaling functions are constructed from direct product
of the 1-D  bases, i.e. 
$\psi_{{\bf j}, {\bf l}}({\bf x})\equiv\psi_{j_1,l_1}(x_1)\psi_{j_1,l_2}(x_2)$, 
and 
$\phi_{{\bf j},{\bf l}}({\bf x})\equiv\phi_{j_1,l_1}(x_1)\phi_{j_1,l_2}(x_2)$,
where ${\bf j}=(j_1,j_2)$ and ${\bf l}=(l_1,l_2)$. For scale $j$, the angular 
scale is 
$2^{2\cdot(10-j)}$ in unit
of a cell. In this paper, we are interested in scales of
$j=8, 7, 6, 5, 4,$ and 3, corresponding to spatial scales of about
0.26, 0.52, 1.04, 2.08, 4.16 and 8.32 h$^{-1}$ Mpc, respectively. 

In doing statistical analysis of the APM-BGC, the effect of the 1,456 holes 
drilled around big bright objects should be taken into account. For instance, 
random samples are generated in the area with the same drilled holes 
as that in the original survey. We also removed a few galaxies placed in the 
holes drilled.

A common problem of statistics for discrete sample is sampling error, which is,
in particular, serious for statistics of non-Gaussianity. Even if the
original matter field is Gaussian, the sampled data must be non-Gaussian
on scales for which the mean number in one cell is small. This is the
non-Gaussianity of shot noise. Any non-Gaussian behavior of the density 
fields will be contaminated by the shot noise. Because of bias might be related
to non-Gaussian process, the detected result is inevitably contaminated by
sampling. Fortunately, the DWT spectrum method is found to be effective for
suppressing the contamination of shot noise. It has been shown that the
non-Gaussianity of shot noise is significant only on scales comparable
with the mean distance of nearest neighbors of objects. Because the wavelet
bases are localized, the central limit theorem guarantees the 
non-Gaussianity of shot noise rapidly and monotonously approaching zero on 
larger scales
(Greiner, Lipa \& Carruthers 1995, Fang \& Pando 1997). Therefore, besides
small scales, the behavior of DWT $j$-spectrum will not be affected by the
shot noise of sampling.

\section{Detection of $j$-spectrum}

\subsection{Two point correlation function}

Before calculating $\tilde{b}_j^{(n)}$, we consider the two-point
angular correlation functions of the ELs and SPs of the APM-BGC.
Because the two-point angular correlation function analysis of
Stromlo-APM redshift survey has been done by Loveday et al. (1995).
One can show the reliability of the DWT method by comparing
our result with the earlier one.

Fig. 2 shows the angular correlation functions, $w_{EL}(\theta)$ and
$w_{SP}(\theta)$, which are obtained from whole sample of the ELs and SPs.  
The dotted lines give the 1 $\sigma$ error calculated from 20 random samples
which are produced by randomizing the positions of the galaxies in the same
field with the same drilled holes as the samples. 

As usual, these correlation functions show a power law, and the amplitude of the
correlation for the ELs is higher than that of the SPs. This is the
well-known segregation of morphology: the clustering of EL galaxies is 
stronger than that of SP (Davis \& Geller 1976.) Actually, this is an 
evidence of 
bias. However, Fig. 2 shows that the ratio $w_{EL}(\theta)/w_{SP}(\theta)$ is 
approximately constant in the angular range of $-0.4 < \log \theta < 0.6$.
Namely, the segregation can be explained by linear bias if $b_{I}$ and
$b_{II}$ are taken to be different constant. In other words, in terms of
two-point correlation functions, no scale-dependent bias is needed,
or at most a very weak scale-dependence. With the DWT analysis, we
are able to measure this scale-dependence more quantitatively.
 
This result of Fig. 2 will not be changed if the error is estimated by 
bootstrap
samples. This is expected, because the number of galaxy pairs is large, the
bootstrap-resampling error is only larger than Poissonian error by a
factor of 1.7 (Mo, Jing \& B\"orner 1992.)

\subsection{Statistics $\tilde{B}^{(2)}_j$ and $\tilde{B}^{(3)}_j$}

Power spectrum is the Fourier counterpart of two-point correlation function.
As mentioned in \S 2.4, the second order statistic 
$\langle \tilde{w}_{j,l}^2 \rangle$ is equivalent to the measure of Fourier 
power spectrum (Pando \& Fang 1998). Therefore, one can expected that the 
second order statistics
$\langle \tilde{b}_{j}^{(2)} \rangle$ should show the same results as that
given by two-point correlation function.

The 2-D extension of the $n=2$ statistic
$\langle \tilde{b}_{j}^{(2)} \rangle$ of Eq.(23) is
\begin{equation}
\tilde{b}_{j_1,j_2}^{(2)}= \frac
{[\sum_{l_1=0}^{2^{j_1}-1}\sum_{l_2=0}^{2^{j_2}-1}
                          \tilde{w}_{{\bf j},{\bf l}}^2]_{I}}
{[\sum_{l_1=0}^{2^{j_1}-1}\sum_{l_2=0}^{2^{j_2}-1}
                          \tilde{w}_{{\bf j},{\bf l}}^2]_{II}}, 
\end{equation}
In order to estimate the error of $\tilde{b}_{j_1,j_2}^{(2)}$, and to avoid
the effect of the drilled holes and boundary, we generated 500 
randomized samples of the EL  and SP, and calculated the average 
$\overline{\tilde{b}_{j,j}^{(2)}}$ and variance from these random samples.
Fig. 3 plots $\tilde{B}^{(2)}_j$ defined as
\begin{equation}
\tilde{B}^{(2)}_j=
\frac{\ \tilde{b}_{j,j}^{(2)}\ }{\overline{\tilde{b}_{j,j}^{(2)}}}.
\end{equation}
The variance from the random samples is also plotted in Fig. 3. 
The $j$-spectrum of $\tilde{B}^{(2)}_j$, indeed, shows flat within 1-$\sigma$, 
only field S2 has a little higher $\tilde{B}^{(2)}_j$ on large scale $j=3$.
This result is in good agreement with the detection of two-point correlation
functions (Loveday et al. 1995.) This shows that the DWT power spectrum 
estimator is reliable (Pando \& Fang 1998.)

It has been emphasized in \S 2.4 that the flatness of the $j$-spectrum
of the second order statistics $\tilde{B}^{(2)}_j$ doesn't imply that
other $j$-spectra will also be flat, especially in the case that the bias 
scale-dependence is caused by non-Gaussian process. Therefore, non-Gaussian 
sensitive statistics are necessary. Fig.4 shows the result of statistics 
$\tilde{B}^{(3)}_j$ which is defined by
\begin{equation}
\tilde{B}^{(3)}_j=
\frac{\ \tilde{b}_{j,j}^{(3)}\ }{\overline{\tilde{b}_{j,j}^{(3)}}},
\end{equation}
where
\begin{equation}
\tilde{b}_{j_1,j_2}^{(3)}= \frac
{[\sum_{l_1=0}^{2^{j_1}-1}\sum_{l_2=0}^{2^{j_2}-1}
                          |\tilde{w}_{{\bf j},{\bf l}}|^3]_{I}}
{[\sum_{l_1=0}^{2^{j_1}-1}\sum_{l_2=0}^{2^{j_2}-1}
                          |\tilde{w}_{{\bf j},{\bf l}}|^3]_{II}}.
\end{equation}
Comparing Fig.4 with Fig.3, it is clear that the $j$-spectra of 
$\tilde{B}^{(3)}_j$ show more deviation from a flat spectrum. Therefore, 
the bias scale-dependence might really originate from non-Gaussian processes.

\subsection{Statistics $A_j$}

A simplest phase-sensitive statistic is the reconstructed distribution 
$\rho^j(x)$ [Eq.(9)], which is given by SFCs. The 2-D extension of
Eq.(9) is
\begin{equation}
\rho^{{\bf j}}({\bf x}) =
  \sum_{l_1= 0}^{2^{j_1} -1}\sum_{l_2= 0}^{2^{j_2} -1}
   w_{{\bf j},{\bf l}} \phi_{{\bf j},{\bf l}}({\bf x}).
\end{equation}
${\bf j}=(j_1,j_2)$ means a reconstruction of the density field on
scale $j_1$ in the dimension $x_1$, and $j_2$ in $x_2$. Therefore,
when ${\bf j}=(j_1,0)$, $\rho^{{\bf j}}({\bf x})$ is a projection
on axis $x_1$, and when ${\bf j}=(0,j_2)$, a projection on axis $x_2$.
In the case of $j_1=j_2=j$, the field is smoothed on the scale 
$j$ in both directions $x_1$ and $x_2$,

A DWT reconstruction on the three fields S1, S2 and S3 was performed.
Since the SFCs, $w_{{\bf j},{\bf l}}$, are proportional to the density
at position ${\bf l}$, the mean density of the reconstructed density 
field is proportional to
\begin{equation}
\overline{w_{{\bf j}}} = \frac{1}{2^{j_1}}\frac{1}{2^{j_2}}
\sum_{l_1=0}^{2^{j_1}-1} \sum_{l_2=0}^{2^{j_2}-1}w_{{\bf j}, {\bf l}}.
\end{equation}
The cells with 
$w_{{\bf j},{\bf l}}>\overline{w_{{\bf j}}}$ are dense regions (clumps),
and cells with $w_{{\bf j},{\bf l}}<\overline{w_{{\bf j}}}$
are underdense regions (voids). Actually, for all ${\bf j}$,
$(\overline{w_{{\bf j}}})_{I}$ and $(\overline{w_{{\bf j}}})_{II}$
are always proportional to the mean number density of galaxies EL and SP,
respectively, and therefore, the ratio
$(\overline{w_{{\bf j}}})_{I}/(\overline{w_{{\bf j}}})_{II}$
doesn't depend on ${\bf j}$.  

Fig.5 shows reconstructed distributions of S1 on scales $j_1=j_2=j=
6,5,4,3$. in which the bold contours denote the area with
$w_{j,l_1;j,l_2}-\overline{w_{{\bf j}}} > 2,3,4...\sigma_j$
from outside to inside successivefully, and the 1 $\sigma$ line
is calculated from 500 random samples for each field.
Fig.5 also shows the underdense regions by light contours corresponding to
$w_{j,i_1;j,l_2}/\overline{w_{{\bf j}}} \leq 1/2,1/8,1/32,1/128 $
successivefully from out to inside. 

A simplified representation of these fields is shown in Fig.6, in which 
the dark area represents the overdense regions,  
$w_{j,l_1;j,l_2} > \overline{w_{{\bf j}}}$, and blank the underdense
regions,  $w_{j,l_1; j,l_2} < \overline{w_{{\bf j}}}$.
The ratio between the clustering strengths of the EL and SP on 
scale $j$ can be measured from Fig.6 by
\begin{equation}
a_j = \frac{({\rm dark\ area})_{EL}}{({\rm dark\ area})_{SP}}
\end{equation}
Similar to the statistic $\tilde{b}^{(2)}_j$, to estimate the error of $a_j$, 
we generated 500 randomized samples of the ELs and SPs, and calculated the
average $\overline{a_j}$ of these random samples. Fig. 7 plots the result of 
statistics $A_j$ defined as
\begin{equation}
A_j=\frac{a_j}{\overline{a_j}}.
\end{equation}
The one $\sigma$ lines of both randomized and bootstrap re-sampling are also
shown in Fig.7. The two error estimates gave about the same confidence level.
The $j$-spectrum of $A_j$ is found to be significantly non-flat. 

\subsection{Statistics $B_j$}

Since $A_j$ is based on statistic $w_{{\bf j},{\bf l}}$, one can expected
that the statistic of $b^{(1)}_j$ of Eq.(21) will give the same conclusion.
The 2-D extension of Eq.(21) is
\begin{equation}
b^{(1)}_{j_1,j_2}= \frac{[\sum_{l_1=0}^{2^{j_1}-1}\sum_{l_2=0}^{2^{j_2}-1}
|w_{{\bf j},{\bf l}}-\overline{w_{{\bf j}}}|]_{I}}
{[\sum_{l_1=0}^{2^{j_1}-1}\sum_{l_2=0}^{2^{j_2}-1}
|w_{{\bf j},{\bf l}}-\overline{w_{{\bf j}}}|]_{II}},
\end{equation}
where we use absolute value 
$|w_{{\bf j},{\bf l}}-\overline{w_{{\bf j}}}|$ because
for odd $n$, 
$\langle (w_{{\bf j},{\bf l}}-\overline{w_{{\bf j}}})^n \rangle$ has
larger relative errors. We calculated $B_j^{(1)}$ defined as
\begin{equation}
B^{(1)}_j=\frac {\ b^{(1)}_{j,j}\ }{\overline{b^{(1)}_{j,j}}},
\end{equation}
where $\overline{b^{(1)}_{j,j}}$ is the average of $b^{(1)}_{j,j}$ over 500
randomized samples of the ELs and SPs. The results of $B^{(1)}_j$ for the
three
fields are given in Fig. 8. The three fields S1, S2 and S3 show the same
feature of the $j$ dependence. The mean SFCs and their variance from the
randomized samples are also plotted in Fig.8. Like the statistics $A_j$,
the spectrum of $B^{(1)}_j$ is non-flat with significance larger than 2
$\sigma$.

Since 2-D sample of galaxies is a $z$ (redshift) projection of their 3-D 
distribution. To study the influence of sample depth on our statistics, we
calculated $B^{(1)}_j$ for samples of ELs and SPs with different 
limit magnitude $m$. Fig.9 plots the result of $B^{(1+)}_j$ which is the 
same as $B^{(1)}_j$ but for sample with limit magnitude $m=16.20$. In 
the case, the numbers of EL and SP galaxies are about 20-25\% less than 
that of $m=16.44$. The $j$-spectra in Fig.9 have completely the same 
features as Fig.8. Therefore, conclusions drawn from Fig. 8 are insensitive 
to the depth of the samples.

In Figs.8 and 9, the $j$-dependence of bias is prominent on scales larger
than 1 h$^{-1}$ Mpc, i.e. larger than the mean distance of nearest
neighbors of galaxies, and therefore, the effect of shot noise is 
negligible.

\section{Discussions and conclusions}

We showed that, instead of one parameter $b$, Eq.(1) introduces a series of
parameters $b^{(n)}_j$ and $\tilde{b}^{(n)}_j$ to describe the bias
scale-dependence. The statistics of $\tilde{b}^{(n)}_j$ with different $n$ 
are independent from
each others. The DWT analysis provides a simple and effective tool of
systematically detecting the scale-dependence of bias parameters on
various orders. This method is effective to be employed for analyzing galaxy
samples which show morphology- and/or luminosity-segregation. 

With this method we detected the $j$-spectra of $\tilde{b}_j^{(n)}$ and 
$b_j^{(n)}$ for the distributions of EL and SP galaxies 
of the APM-BGC samples. The general results indicate that for second order
statistics, i.e. two-point correlation function and power spectrum, the
bias is approximately scale-independent, but not so for higher order or
phase-sensitive (non-Gaussian) statistics. The result is consistent
with the following fact: most evidences for weak scale dependence of bias
are from statistics of the two-point correlation functions and power
spectrum (Kauffmann, Nusser \& Steinmetz 1997), while the evidence for
scale-dependence is from phase-sensitive statistics (Sigad et al. 1998).
Therefore, the scale-dependence of galaxy bias may have the same origin
as the non-Gaussianity of galaxy distribution. Linear evolution
cannot cause non-Gaussianity of mass distributions if the initial 
perturbations are Gaussian. The scale-dependence of bias of
galaxy distribution is most unlikely due to the non-linear evolution
of gravitational clustering, and the non-local relationship between galaxy 
formation and their environment.

With this in mind, the information of bias scale-dependence is worth for 
developing models of galaxy formation. Indeed, despite the current 
detection of bias 
scale dependence is still very preliminary, the result is already able to
set useful constraint on models of hierarchical clustering.  
In these models, galaxy correlations are generally assumed 
to be described by the hierarchical relation 
$\xi_n = Q_n \xi_2^{n-1}$ where $\xi_n$ is the $n$-th
order correlation function, and $Q_n$ are constants (White 1979).
If these hierarchical relations hold exactly, the second order (two-point)
correlation function plus all constants $Q_n$ (which may be different for
different types of galaxies) completely characterize the clustering of 
galaxies, including their higher order correlations. This is, all
$\tilde{B}^{(n)}_j$
and $B^{(n)}_j$ can be represented by second order correlation function 
plus all (scale-independent) constants $Q_n$. Hence, if bias is 
scale-independent on second order, it will be scale-independent on all
orders. Therefore, the non-flatness of $j$-spectra of
$\tilde{B}^{(n)}_j$ and $B^{(n)}_j$ for sample APM-BGC implies that
the hierarchical relations may not hold exactly, or the coefficients, $Q_n$,
are scale-dependent. Similar conclusion has also been drawn from
the detection of the scale-scale correlations of the Ly$\alpha$ forests
of QSO absorption spectrum (Pando et al. 1998.) Thus, the detection of
$\tilde{B}^{(n)}_j$ and $B^{(n)}_j$, joining with other higher order 
statistics, is effective to reveal the details of the hierarchical
clustering scenario.

We thank Drs. J. Pando and Y.P. Jing for many helpful comments. ZGD 
and XXY were supported by the National Science Foundation of China.

\clearpage

\begin{figure}
\epsscale{1.0} \plotone{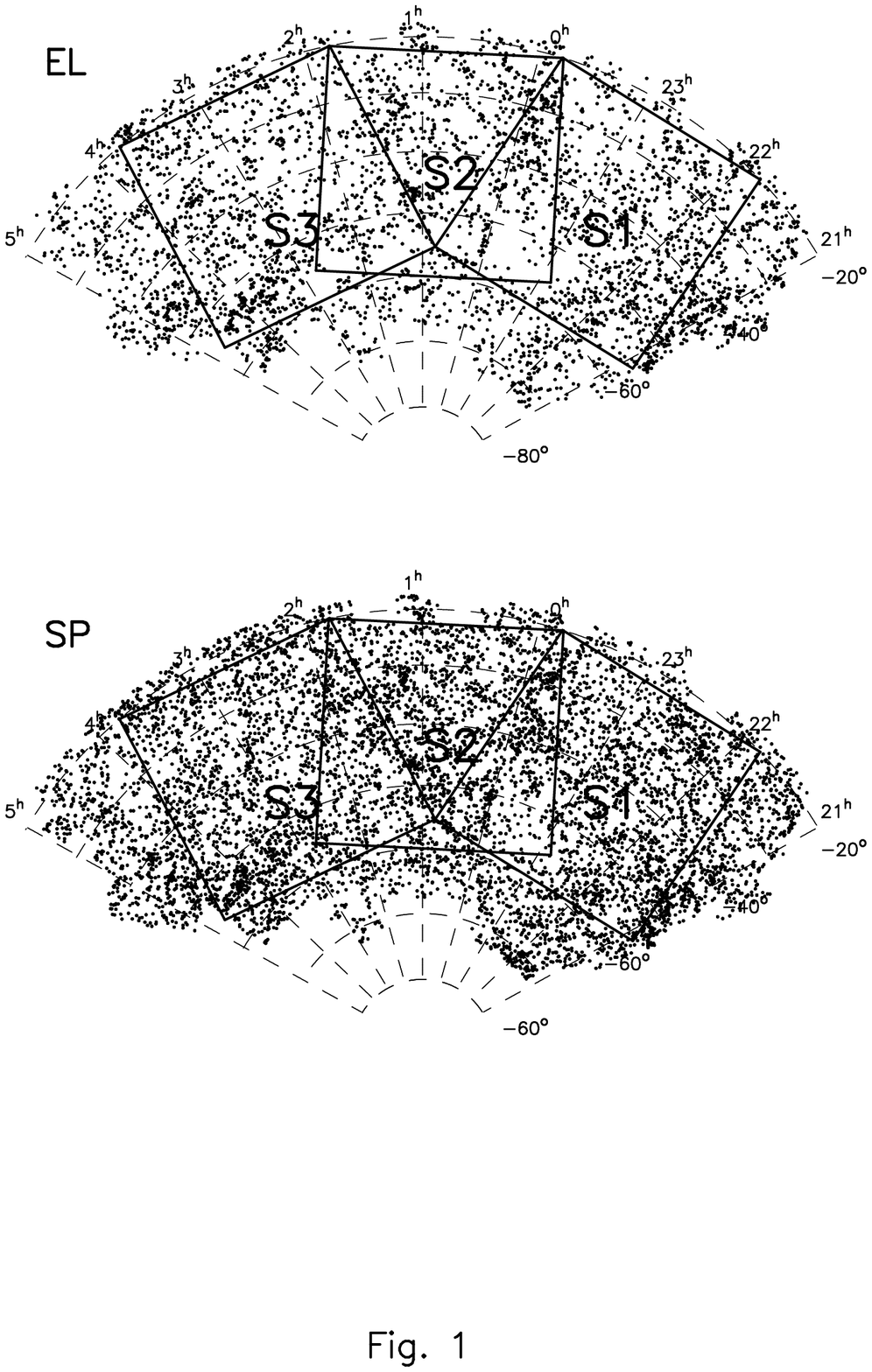}
\figcaption{The equal project of the APM bright galaxy distribution 
and three regions S1, S2 and S3, in which the wavelet analysis has been 
done. The top and bottom panels are for EL and SP galaxies respectively.
  }
\label{fig1}
\end{figure}

\begin{figure}
\epsscale{1.0} \plotone{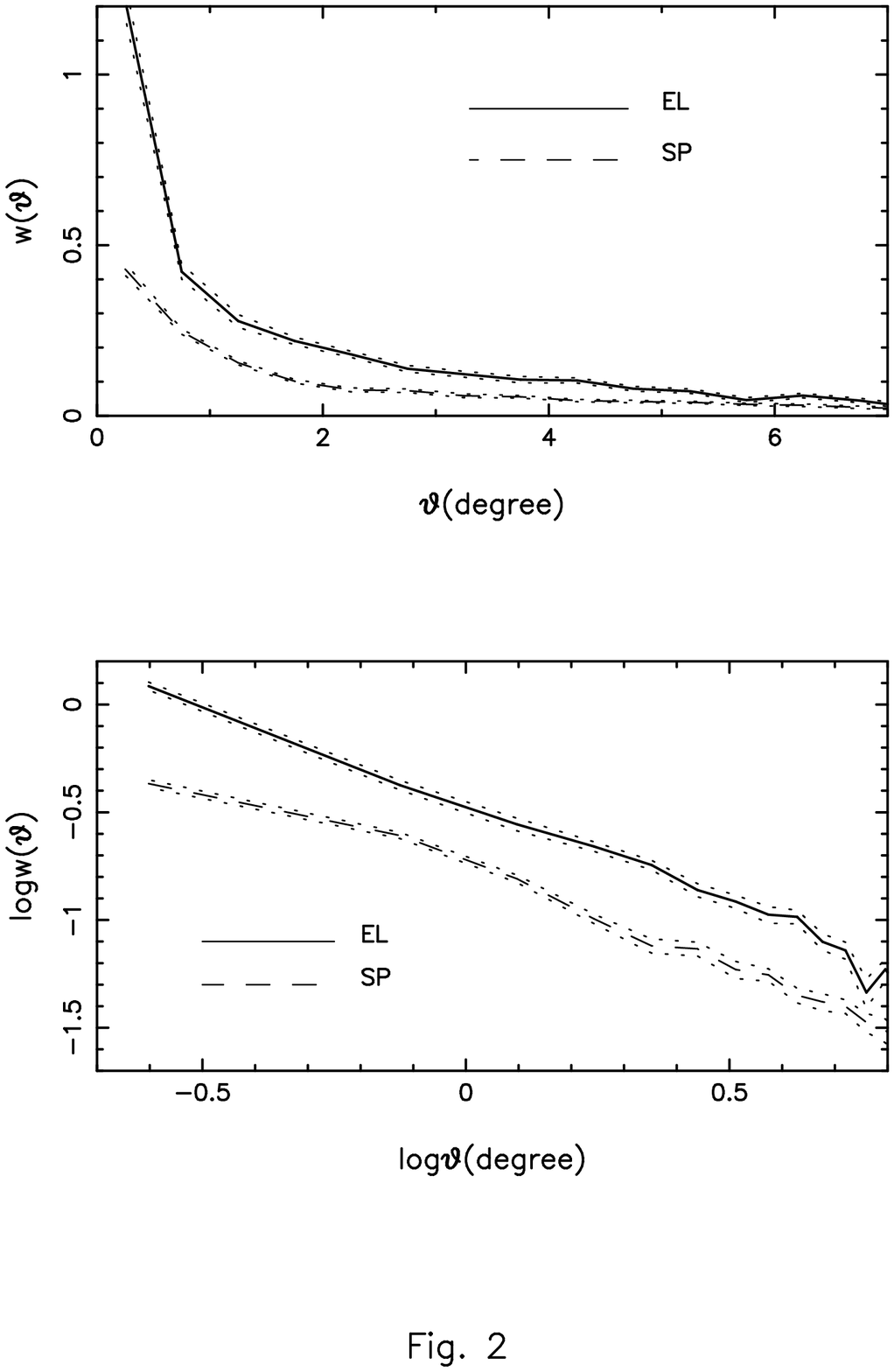}
\figcaption{Two point angular correlation functions of EL (solid lines) 
and SP (dashed lines) galaxies in APM bright galaxy catalogue.
\label{fig2}}
\end{figure}

\begin{figure}
\epsscale{1.0} \plotone{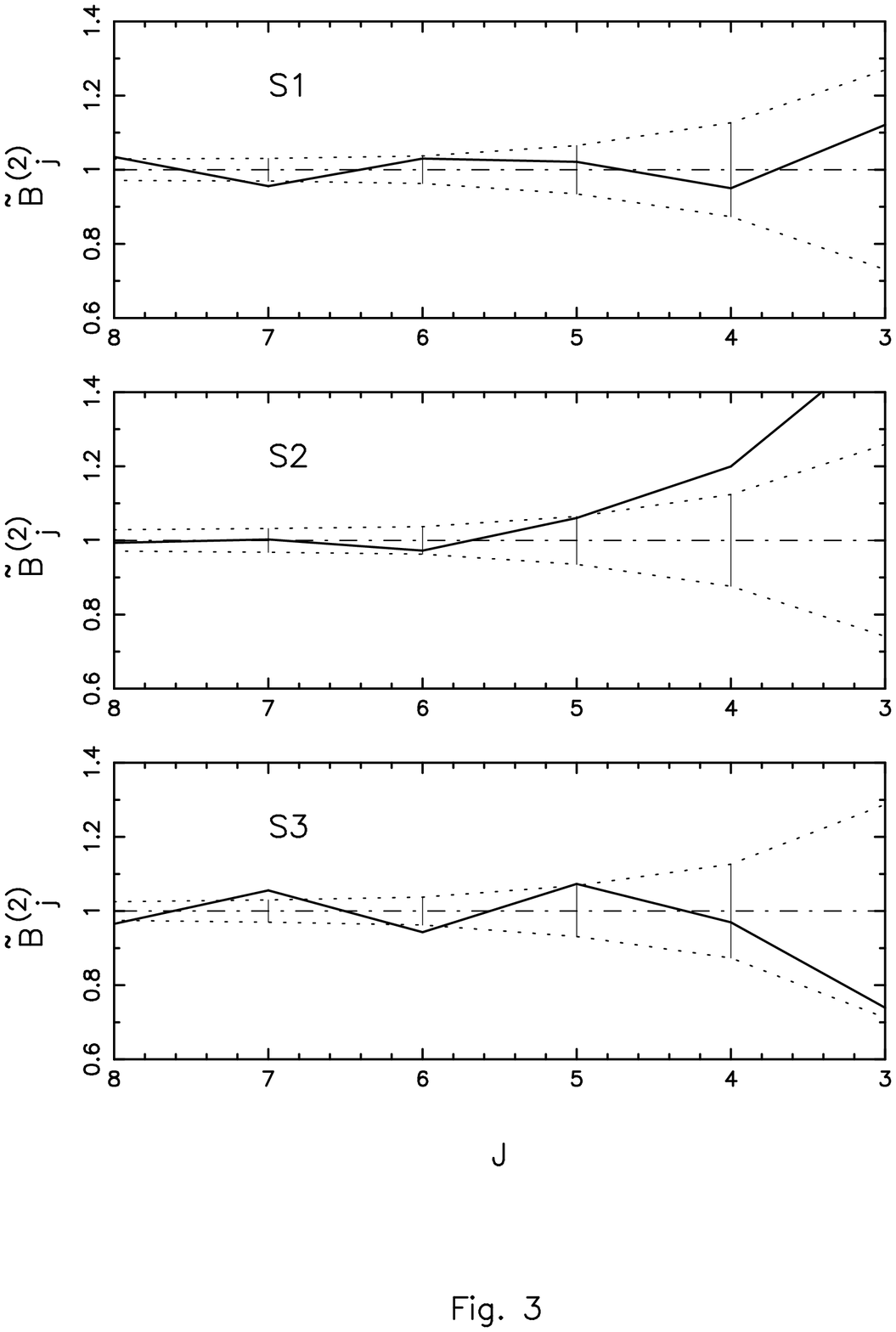}
\figcaption{$\tilde{B}_{j}^{(2)}$ vs. j for three fields S1, S2 and S3. 
The  mean of 
$\tilde{B}_{j}^{(2)}$ and 1 $\sigma$ error given by randomized samples are 
plotted  by dot-dashed and dashed lines, respectively.
\label{fig3}}
\end{figure}

\begin{figure}
\epsscale{1.0} \plotone{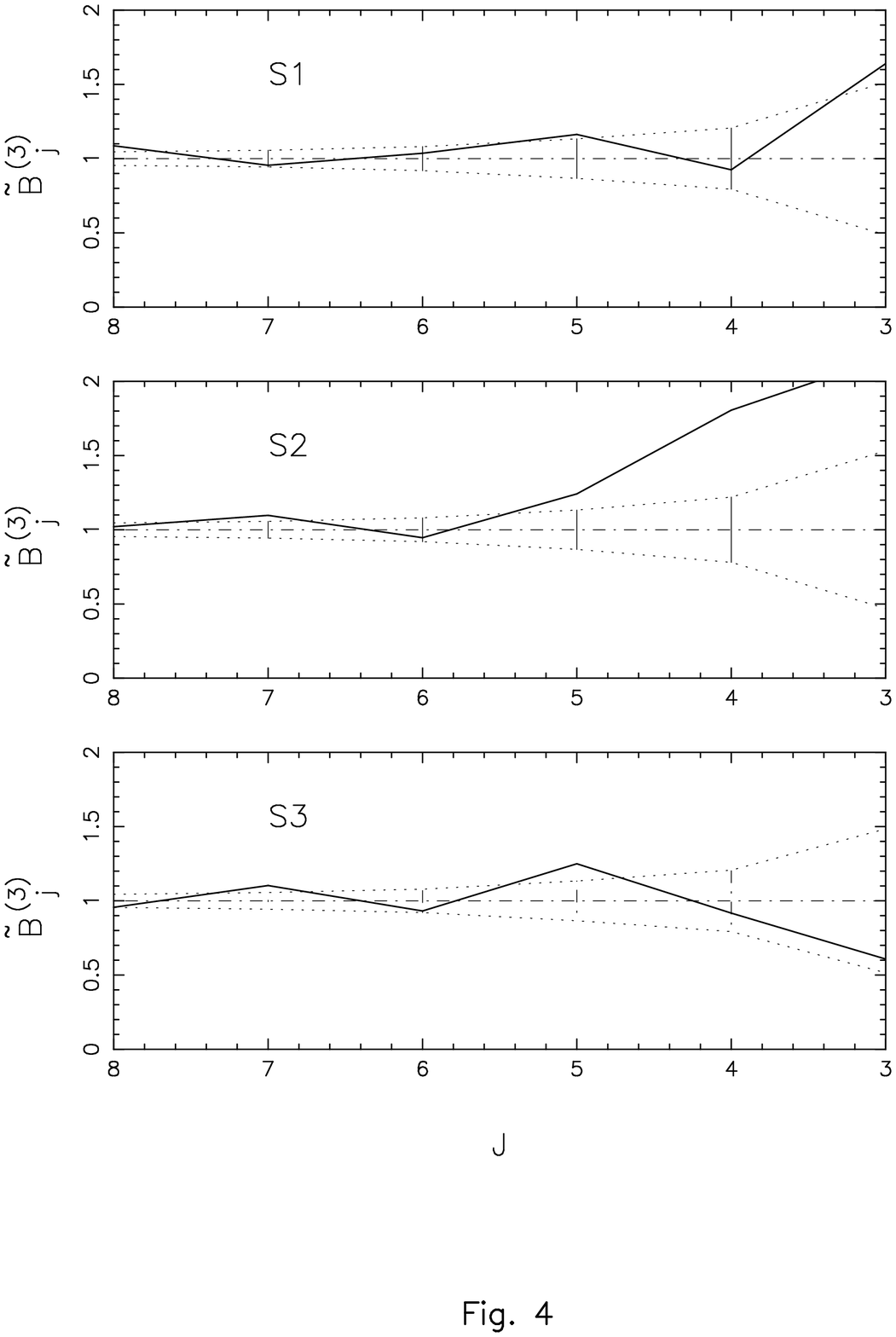}
\figcaption{$\tilde{B}_{j}^{(3)}$ vs. j for three fields S1, S2 and S3. The 
mean of $\tilde{B}_{j}^{(3)}$ and 1 $\sigma$ error given by randomized 
samples are plotted by dot-dashed and dashed lines, respectively.
\label{fig4}}
\end{figure}

\begin{figure}
\epsscale{1.0} \plotone{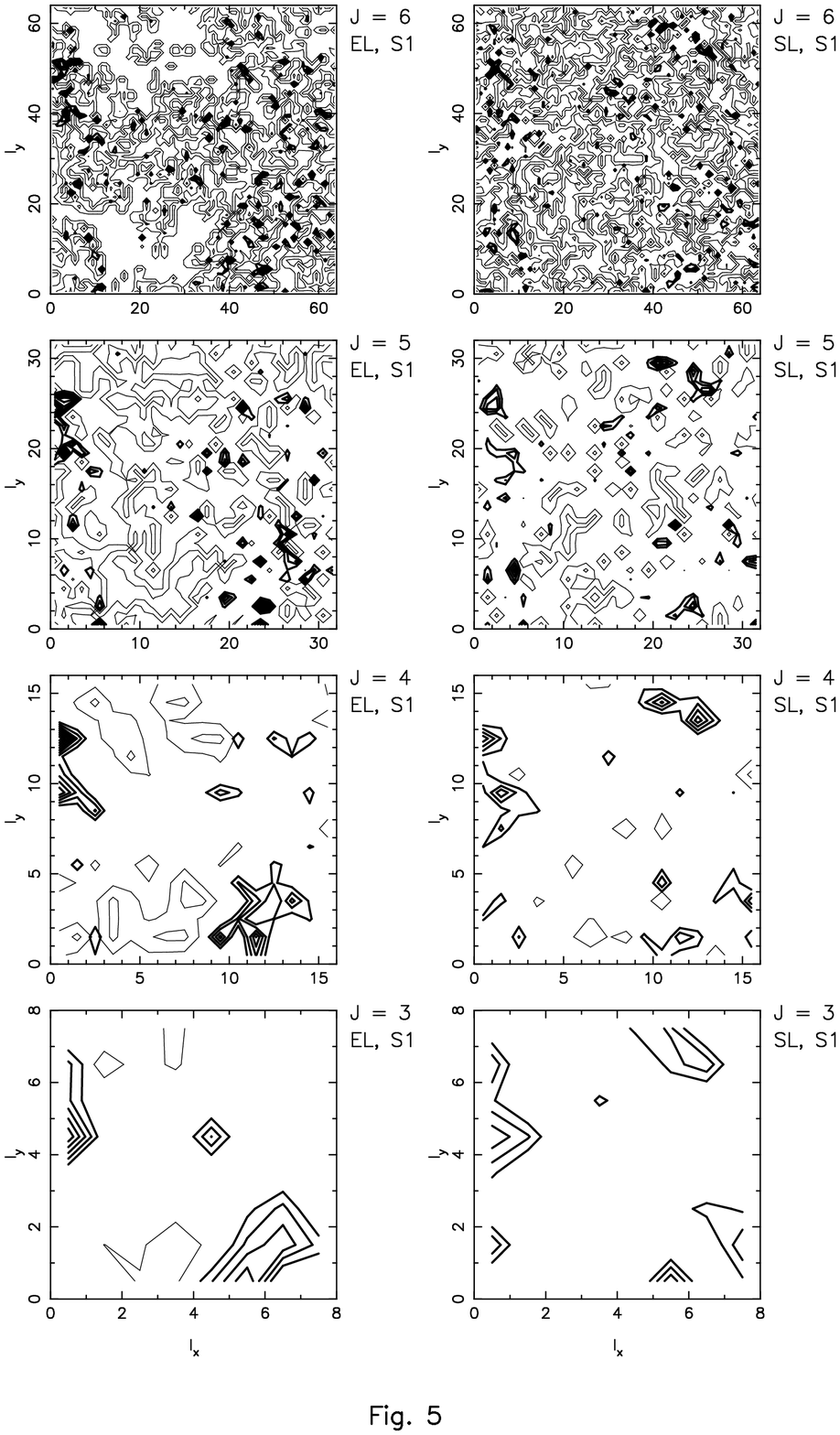}
\figcaption{Reconstructed distributions of S1 on scales $j_1=j_2=6,5,4,3$
from top to bottom. Left side results are for EL galaxies and right for SPs.
The bold contours are for 
$w_{j,l_1;j,l_2}-\overline{w_{{\bf j}}} > 2,3,4...\sigma_j$
from outside to inside successivefully, and the light contours denote 
the area with 
$w_{j,i_1;j,l_2}/\overline{w_{{\bf j}}} \leq$ 1/2,1/8,1/32,1/128
from outside to inside successivefully.
\label{fig5}}
\end{figure}

\begin{figure}
\epsscale{1.0} \plotone{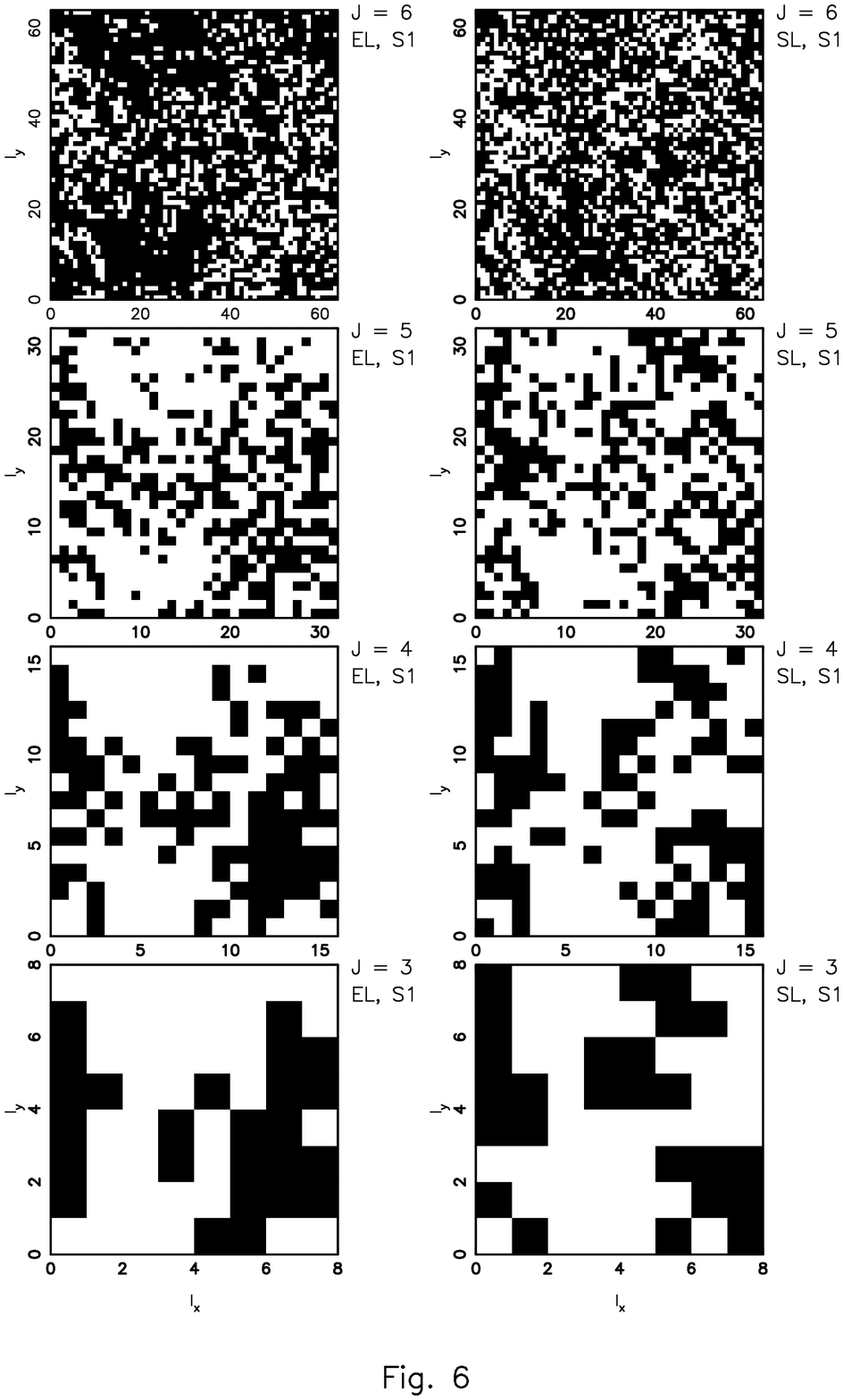}
\figcaption{A simplified representation of Fig. 5. Dark area denote the
overdense regions, $w_{j,l_1;j,l_2} > \overline{w_{j}}$, and the blank area
the underdense regions, $w_{j,l_1; j,l_2} < \overline{w_{j}}$, where $j_1=j_2=j$.
\label{fig6}}
\end{figure}

\begin{figure}
\epsscale{1.0} \plotone{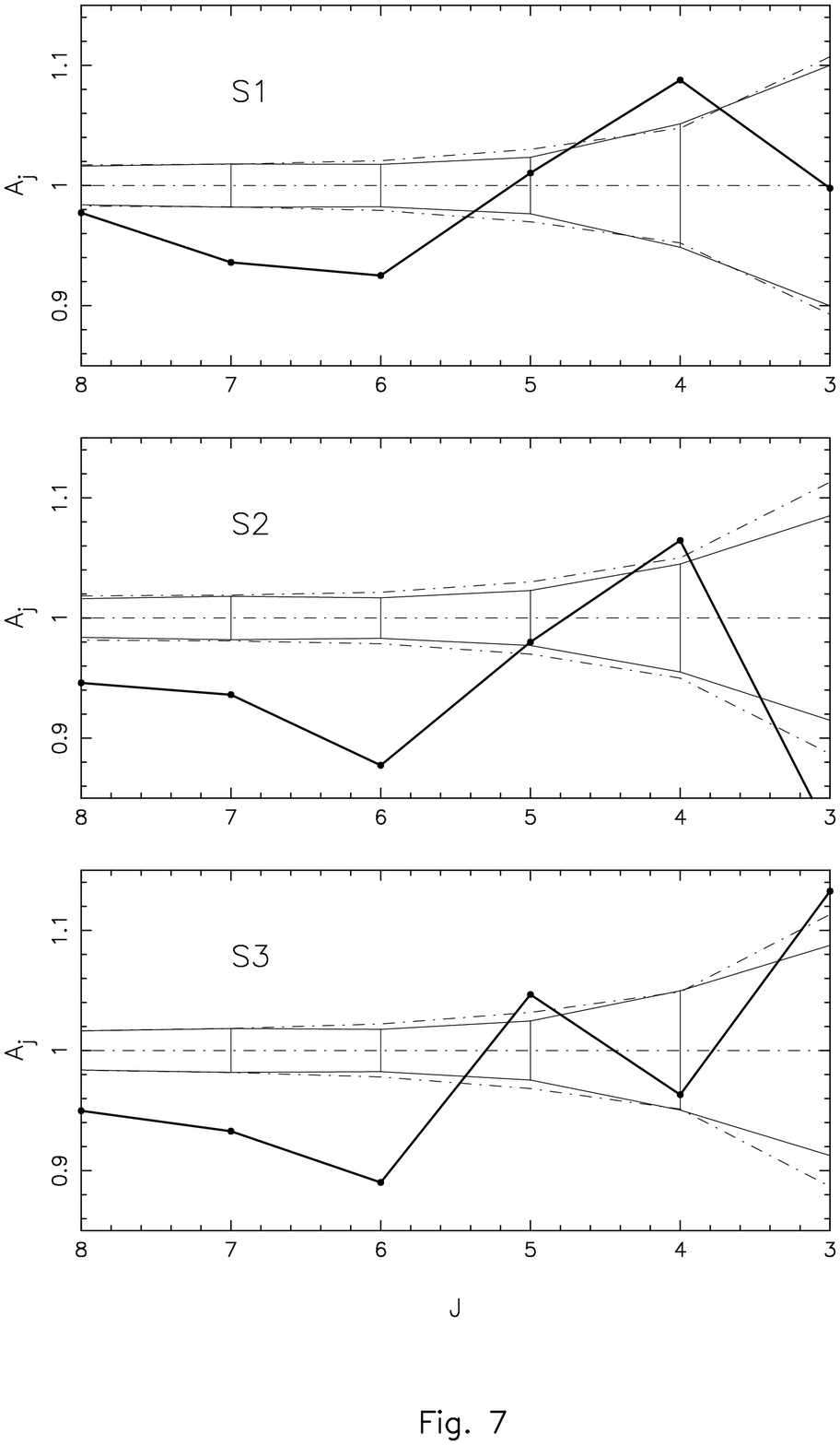}
\figcaption{$A_{j}$ vs. j for three fields (solid lines). It shows the mean 
of $A_{j}$ (thick solid lines), and 1 $\sigma$ errors given by randomized samples 
(dotted lines) and bootstrap-resampling (thin solided lines.)
\label{fig7}}
\end{figure}

\begin{figure}
\epsscale{1.0} \plotone{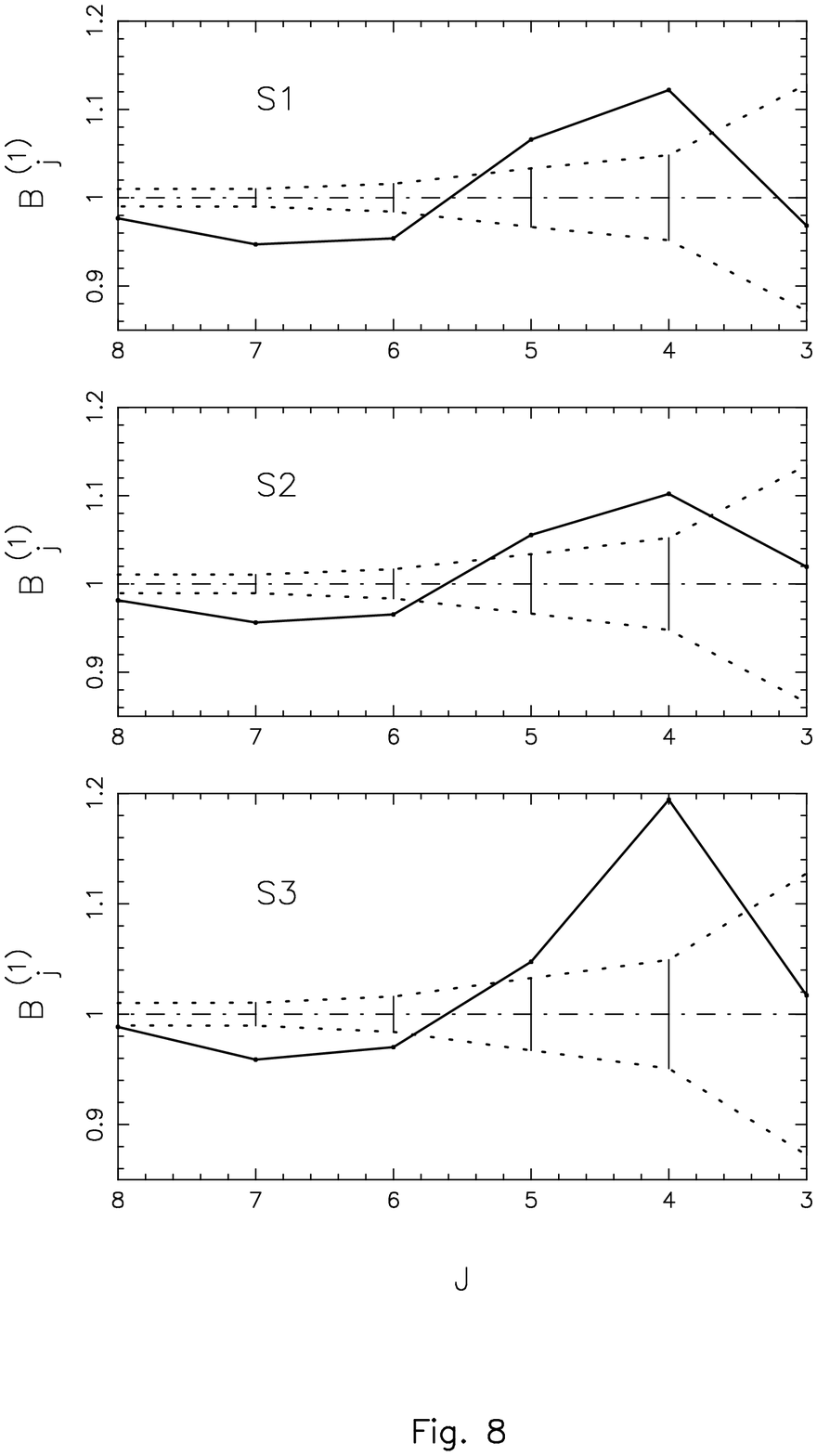}
\figcaption{$B_{j}^{(1)}$ vs. j for three fields (solid lines.)  The mean of 
$B_{j}^{(1)}$ and 1 $\sigma$ error given by randomized samples are plotted 
by dot-dashed and dashed lines, respectively.
\label{fig8}}
\end{figure}

\begin{figure}
\epsscale{1.0} \plotone{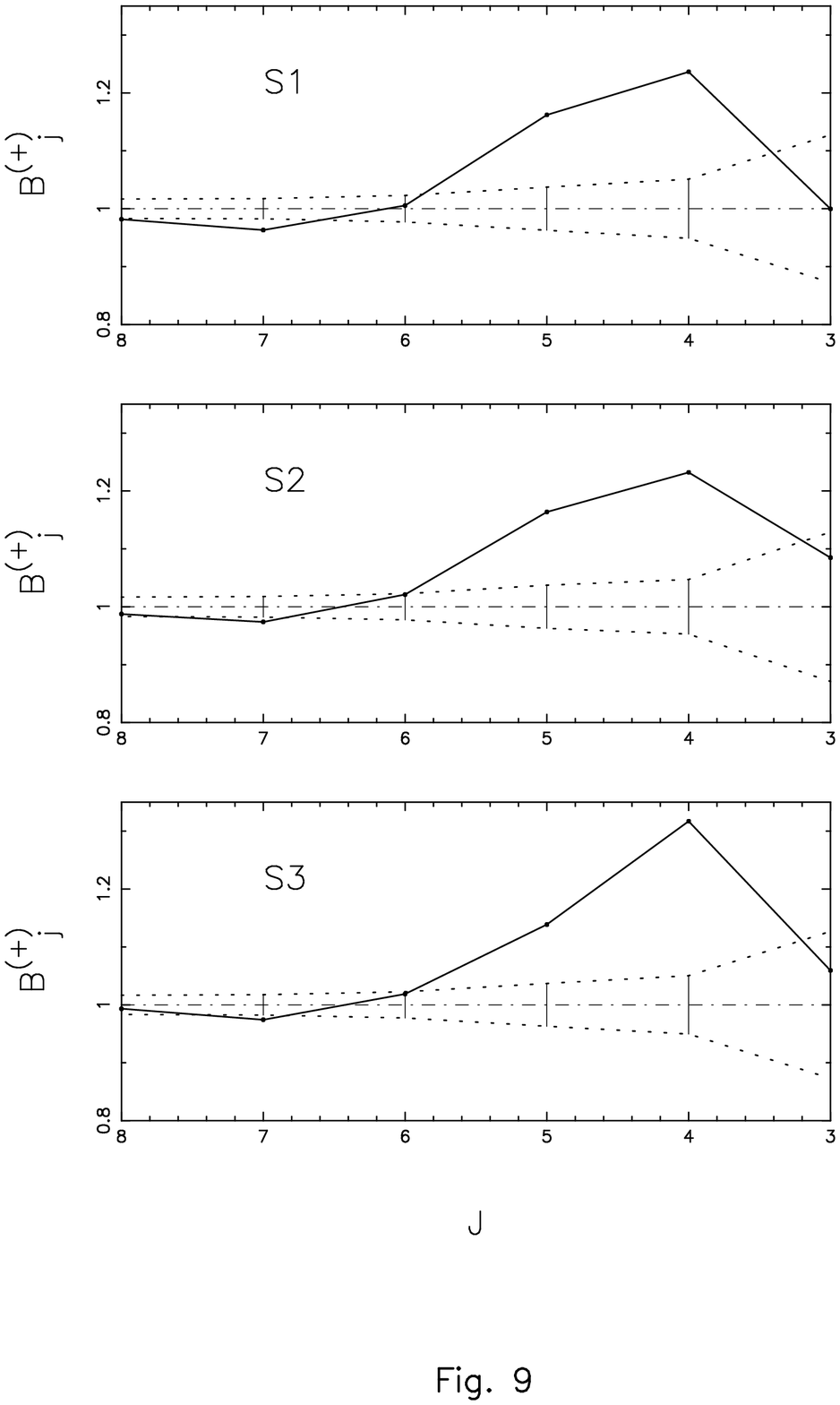}
\figcaption{B$^{(1+)}_j$ vs. j for three fields (solid lines.) The mean of 
B$^{(1+)}_j$ and 1 $\sigma_j$ error given by randomized samples are plotted 
by dot-dashed and dashed lines, respectively.
\label{fig9}}
\end{figure}

\newpage

\begin{references}
\reference{}Adler, R.J. 1981, the Geometry of Random Fields, Wiley,
New York.
\reference{} Bower, R.G., Coles, P., Frenk, C.S. and White, S.D.M.
1993, \apj, 405, 403
\reference{ } Catelan, P., Coles, P., Matarreses, S. and Moscardini, L.
1994, \mnras, 268, 966
\reference{ } Coles, P. 1993, \mnras, 262, 1065.
\reference{ } Couchman, H.M.P. and Carlberg, R.G. 1992, \apj, 389, 453
\reference{daubechies} Daubechies, I. 1992, Ten Lectures on Wavelets,
SIAM, Philadelphia
\reference{} Davis, M. and Geller, M.J.. 1976, \apj, 208, 13.
\reference{ } Deng, Z.G., Xia, X.Y., Fang, L.Z. and B\"orner, G. 1997,
Astrophysical Report, No.2,  82
\reference{ } Dekel, A., Burstein, D. and White, S.D.M. 1996, 
astro-ph/9611108
\reference{ } Dekel, A. and Rees, M. 1987, Nature, 326, 455
\reference{} Dressler, A. 1980, \apj, 236, 351
\reference{} Fang, L.Z. and Pando, J., 1997, in 5th Current
Topics of Astrofundamental Physics, eds. N. Sanchez and A. Zichchi,
World Scientific, Singapore, 616
\reference{} Greiner, M., Lipa, P. \& Carruthers, P. 1995,
  Phys. Rev. E, 51, 1948
\reference{ } Kaiser, N. 1984, \apj, 284, L9
\reference{ } Kauffmann, G., Nusser, A. and Steinmetz, M. 1997, \mnras, 286, 795
\reference{} Loveday, L, Peterson, B. A., Efstathiou, G, and Maddox S. J., 
1992, \apj, 390, 338
\reference{} Loveday, L, 1996, \mnras, 278, 1025
\reference{} Loveday, L., Maddox, S.J., Efstathiou, G. \& Peterson, B.A. 
             1995, \apj, 442, 457
\reference{ } Meyer, Y. 1993, Wavelets: Algorithms and Applications, SIAM, 
    Philadelphia
\reference{} Mo, H.J., Jing, Y.P. \& B\"orner, G. 1992, \apj, 392, 452
\reference{} Mo, H.J., Jing, Y.P. and White, S.D.M. 1996, \mnras, 284, 189
\reference{} Pando, J. and Fang, L.Z., 1996, \apj, 459, 1
\reference{} Pando, J. and Fang, L.Z., 1998, Phys. Rev. E57, 3593
\reference{} Pando, J., Lipa, P., Greiner, M. and Fang, L.Z., 1998, \apj, 469, 9
\reference{} Peebles, P. J. E., 1980, The Large Scale Structure of the
Universe, Princeton, NJ., Princeton Univ. Press.
\reference{} Rees, M. 1985, \mnras, 213, 75p
\reference{} Saunders, W., Roiwan-Robinson, M., \& Lawrence, A. 1992,
\mnras, 258, 134
\reference{} Sigad, Elder, A., Dekel, A., Strauss, M.A. \& Yahil, A. 1998 
\apj, 495, 516
\reference{} Staruss, M.A., Davis, M., Yahil, A. and Huchra, J.P. 1992,
\apj, 385, 421
\reference{van} Vanmarcke, E. 1983, Random Field, MIT Press.
\reference{} Xia, X.Y., Deng, Z.G and Zhou, Y.Y. 1987, in Observational 
Cosmology, eds. Hewitt, A., Burbidge and L.Z.Fang, Reidel Pub. Co. 363
\reference{yamada}Yamada, M. and Ohkitani, K. 1991, Prog. Theor. Phys.,
86, 799
\reference{}White, S.D.M. 1979, \mnras, 186, 145
\end{references}
\end{document}